\documentstyle[floats,prb,aps]{revtex}
\input epsf

\begin{document}
\draft

\smallskip
%%%%%%%%%%%%%%%%%%%%%%%%%%%%%%%%%%%%%%%%%%%%%%%%%%%%%%%%%%%%%%%%%%%%%%%%%%
%                               Title                                    %
%%%%%%%%%%%%%%%%%%%%%%%%%%%%%%%%%%%%%%%%%%%%%%%%%%%%%%%%%%%%%%%%%%%%%%%%%%
\twocolumn[\hsize\textwidth\columnwidth\hsize\csname @twocolumnfalse\endcsname
\title{Nesting Induced Precursor Effects: a Renormalization Group Approach}
\author{ F. Vistulo de Abreu$^{a}$ and Benoit Dou\c{c}ot$^{b}$}
\address{$^{a}$Departamento de F\'{\i}sica, Universidade de Aveiro,
3810 Aveiro, Portugal.}
\address{$^{b}$Laboratoire de Physique Th\'{e}orique et Hautes Energies,
Jussieu, 75252 Paris Cedex 05}
\date{\today}
\maketitle
\widetext
\begin{abstract}
\leftskip 54.8pt \rightskip 54.8pt We develop a controlled weak
coupling renormalization group (RG) approach to itinerant
electrons. Within this formalism we rederive the phase diagram for
two-dimensional (2D) non-nested systems. Then we study how nesting
modifies this phase diagram. We show that competition between p-p
and p-h channels, leads to the manifestation of unstable precursor
fixed points in the RG flow. This effect should be experimentally
measurable, and may be relevant for an explanation of pseudogaps
in the high temperature superconductors (HTC), as a crossover
phenomenon.
\end{abstract}

 ]
%%%%%%%%%%%%%%%%%%%%%%%%%%%%%%%%%%%%%%%%%%%%%%%%%%%%%%%%%%%%%%%%%%%%%%%%%%
%                              body of paper                             %
%%%%%%%%%%%%%%%%%%%%%%%%%%%%%%%%%%%%%%%%%%%%%%%%%%%%%%%%%%%%%%%%%%%%%%%%%%
\narrowtext

\smallskip An important clue to understand the physics of the HTC may come
from the unconventional normal state properties preceding
superconducting ordering: a pseudogap is opened, which manifests
through an apparent continuous reduction in the density of
states\cite{ARPES} at a temperature $T^{*}$, higher than the
superconducting critical temperature, $T_{c}$. The precise
definition of this incipient ordering mechanism is still subject
of discussion\cite{Tallon2000}$^{,}$\cite {Timusk99}. Two main
pictures have been proposed depending on whether the correlations
being built up in the pseudogap are of superconducting\cite
{Randeria} or (spin, or charge) density wave nature\cite{RVB}.
However no experimental conclusive proof has been given so far
favoring any of these pictures.

In the former, it is assumed that pre-formed pairs appear at the
pseudogap energy scale $T^{*}$, but coherence would only be established at $%
T_{c}.$ Some support in favor of this scenario is provided for
instance by ARPES measurements. They indicate that the pseudogap
shares some of the superconducting properties like the gap
symmetry. Furthermore, they reveal the presence of a non-vanishing
gap in the vicinity of the $\left( \pi ,0\right) $ point already
above  $T_{c}$\cite{ARPES2}. Nevertheless, it has been argued that
NMR and heat capacity data show that superconductivity and
pseudogap are competing\cite{Loram94}, like antagonistic phases.
Also, $T^{*}$ merges with $T_{c}$ at a critical doping in the
overdoped superconducting region, when the condensation energy,
the critical currents and the superfluid density reach sharp
maxima. This has been argued to be irreconcilable with a preformed
pairing scenario\cite {Tallon2000}.

The amplitude of incommensurate peaks observed in inelastic
neutron scattering varies upon doping in a way highly related to
the pseudogap and falls to zero in the overdoped
region\cite{Tallon2000}. ARPES has clearly shown the existence of
a Fermi surface which is lying not too far from the diagonals of
the first Brillouin zone\cite{ARPES2}, therefore providing an
approximate nesting.

In the ideal case of perfect nesting, it is well known that
density wave and superconducting instabilities are strongly
interfering\cite {Solyom79}$^{-}$ \cite{Balents}. In this work we
present a simple one-loop RG approach to electrons on 2D Fermi
surfaces with flat nested regions, away from half-filling. We show
that the strong competition arising from particle-particle and
particle-hole corrections when nesting of the Fermi surface (FS)
is included, naturally induces instabilities that can be preceded
by other incipient instabilities. In the RG sense this appears as
a {\it crossover} phenomenon: the RG flow passes near an unstable
fixed point, capturing the physics of this fixed point at
intermediate energies, before the system orders in the final
superconducting state. This scenario naturally provides two energy
scales and therefore bears striking similarities with the
pseudogap phase of the HTC's. The condition of perfect nesting may
be relaxed provided interactions are large enough.
%%%%%%%%%%%%%%%%%%%%%
%      figure       %
%%%%%%%%%%%%%%%%%%%%%
\begin{figure}[tbp]
     \begin{center}
       \leavevmode
       \hbox{%
       \epsfxsize 2.1in \epsfbox{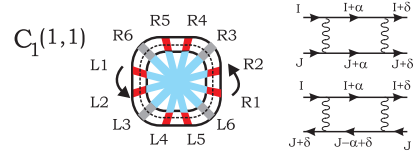}}
       \end{center}
\caption{ Left: Tomographic parameterisation of a Fermi surface
with 3 patches per side ($N_{c}=3$): 2 on nested regions (N=2) and
1 in curved portions (M=1). The $C_{1}\left( 1,1\right) $ is
illustrated. Right: Diagrams considered.}
\end{figure}
%%%%%%%%%%%%%%%%%%%%%
%      end          %
%%%%%%%%%%%%%%%%%%%%%

We parametrize a 2D Fermi surface with a set of $4N_{c}$ patches,
as illustrated in figure 1. The kinetic energy dispersion is
linearized around the Fermi surface, each patch being assigned a
Fermi velocity, $v_{{I}}.$ In this work all $v_{{I}}$ are assumed
equal as the results do not depend significantly on smooth Fermi
velocity modulations. The most general spin rotational invariant
interaction term which remains in an effective low energy theory
only involves incoming electrons on opposite sides of the Fermi
surface\cite{Douçot,Shankar}. This remark enables us to separate
the set of patches in two complementary subsets called right ($R$)
and left ($L$) patches, for convenience. We may therefore write:

\begin{eqnarray}
{\cal H}_{C}
&=&\frac{1}{2N_{c}L}\sum_{{I,K,\delta}}\,\sum_{q\simeq 0}\
(C_{{\delta }}^{c}({I,J})\,J_{q,{\delta }}^{R}\left( {I}%
\right) \,J_{-q,-{\delta }}^{L}\left( {K}\right) \nonumber
\\ &&+C_{{\delta }}^{s}({I,J})\,{\bf J}_{q,{\delta
}}^{R}\left( {I}\right) \cdot {\bf J}_{-q,-{\delta }}^{L}\left( {
K}\right) ) \label{interaction}
\end{eqnarray}
where charge $J_{{\delta }}^{R}({I})=\sum_{\tau ,\tau ^{\prime }}:R_{%
{I+\delta },\tau }^{+}\;\delta _{\tau ,\tau ^{\prime }}\,\,R_{{I}%
,\tau ^{\prime }}:$ , and spin currents ${\bf J}_{{\bf \delta }}^{R}({I}%
)=\sum_{\tau ,\tau ^{\prime }}:R_{{I+\delta },\tau }^{+}\;{\bf \sigma }%
_{\tau ,\tau ^{\prime }}\,\;R_{{I},\tau ^{\prime }}:,$ have been
used, and momentum variables within each patch have been omitted
to simplify the notation. Here chain indices could have been
written in bold as they could represent vectors\cite{Fernao}. In
that case we could, for instance, consider a bilayer of 2D
electron systems: the present formalism is in fact quite general
allowing the treatment of many different systems. Eq. $\left(\ref
{interaction}\right) $ allows an extremely simple derivation of
one-loop equations: only the two diagrams in fig. 1, need to be
considered. Requiring the invariance of the two-particle vertex
under cutoff reduction we arrive at ($
\partial \equiv
\partial /\partial \ln D$ and $C_{i}\equiv C_{i}/2\pi N_{c} $) :
\begin{eqnarray}
\partial C_{{\delta}}^{c}({I,K})\text{{}}&=&\text{{}}\sum_{{%
\alpha}}\delta ^{c}({I,K,\delta,\alpha})\,{\Large (}C_{{\alpha }%
}^{c}({I,K})C_{{\delta-\alpha}}^{c}({I+\alpha,K+\alpha}) \nonumber
\\
&&\hspace{-1.7cm}+3C_{{\alpha}}^{s}({I,K})C_{{\delta-\alpha}%
}^{s}({I+\alpha,K+\alpha}){\Large)/} \left(v_{{I+\alpha}%
}+v_{{K+\alpha}}\right)   \nonumber \\
&&\hspace{-1.7cm}-\sum_{{\alpha}}\delta ^{ZS}({I,K,\delta,\alpha})%
{\Large (}C_{{\alpha}}^{c}({I,K-\alpha+\delta})C_{{\delta -\alpha
}}^{c}({I+\alpha,K})  \label{RG1} \\
&&\hspace{-1.9cm}+3C_{{\alpha}}^{s}({I,K-\alpha+\delta})\,C_{{%
\delta-\alpha}}^{s}({I+\alpha,K}){\Large)/} \left( v_{{%
I+\alpha}}+v_{{K-\alpha +\delta}}\right)   \nonumber
\end{eqnarray}
for charge couplings, whereas for spin couplings we get:

\begin{eqnarray}
\partial C_{{ \delta }}^{s}({ I,K}) &=&\sum_{{ \alpha }}\delta ^{c}(%
{ I,K,\delta ,\alpha })\,{\Large (}C_{{ \alpha }}^{s}({ I,K})C_{%
{ \delta -\alpha }}^{c}({ I+\alpha ,K-\alpha })  \nonumber \\
&&\hspace{-0.7cm}+C_{{ \alpha }}^{c}({ I,K})C_{{ \delta -\alpha }%
}^{s}({ I+\alpha ,K-\alpha })  \label{RG2} \\
&&\hspace{-1.7cm}-2C_{{ \alpha }}^{s}({ I,K})\,C_{{ \delta -\alpha }%
}^{s}({ I+\alpha ,K-\alpha }){\Large )}/ \left( v_{{ I+\alpha }%
}+v_{{ K+\alpha }}\right) -  \nonumber \\
&&\hspace{-1.7cm}-\sum_{{ \alpha }}\delta ^{ZS}({ I,K,\delta ,\alpha })%
{\Large (}C_{{ \alpha }}^{c}({ I,K-\alpha +\delta })C_{{ \delta
-\alpha }}^{s}({ I+\alpha ,K})  \nonumber \\
&&\hspace{-0.7cm}+C_{{ \alpha }}^{s}({ I,K-\alpha +\delta })C_{{ %
\delta -\alpha }}^{c}({ I+\alpha ,K})  \nonumber \\
&&\hspace{-1.7cm}+2C_{{ \alpha }}^{s}({ I,K-\alpha +\delta })\,C_{{ %
\delta -\alpha }}^{s}({ I+\alpha ,K}){\Large )}/ \left( v_{{ %
I+\alpha }}+v_{{ K-\alpha +\delta }}\right)   \nonumber
\end{eqnarray}
Here $\delta ^{c}$ and $\delta ^{ZS}$ implement phase space
restrictions: they are equal to 1 iff all electrons involved in
the scattering are kept on the FS along the process. This happens
if the electrons are on nested regions of the FS, or, in the case
of the Cooper channel, if ${ I=K}$. Otherwise $\delta ^{c}$ and
$\delta ^{ZS}$ are set equal to zero.

In order to characterize the possible instabilities in the system, we
introduce \ several response functions, of the form: ${\cal R}_{{ \delta }%
}\left( { I,K}\right) =$ $-i\int dt\,e^{i\omega t}$ $\left\langle
T\left\{ {\cal O}_{{ \delta }}\left( { I;}t\right) \,{\cal O}_{{ %
\delta }}^{+}\left( { K;}0\right) \right\} \right\rangle .$ For
singlet superconductivity we denote ${\cal R}_{{ \delta }}\left( {
I,K}\right) \equiv S_{{ \delta }}^{S}\left( { I,K}\right) $ and
the order
parameter is: ${\cal O}_{SSC,{ \delta }}\left( { I;}t\right) =\frac{1}{%
\sqrt{2}}\sum_{\tau }\tau \,R_{{ I+\delta },\tau }\left( t\right) L_{{ %
I},-\tau }\left( t\right) .$ For triplet superconductivity, ${\cal R}_{{ %
\delta }}\left( { I,K}\right) \equiv S_{{ \delta }}^{T}\left( { I,K}%
\right) ,$ and ${\cal O}_{TSC,{ \delta }}\left( { I;}t\right) =\frac{1%
}{\sqrt{2}}\sum_{\tau }R_{{ I+\delta },\tau }\left( t\right) L_{{ I}%
,-\tau }\left( t\right) .$ As interactions do not connect pairs of
particles with different ${ \delta }$ (because of momentum
conservation)\cite{Douçot}, the problem of finding the most
divergent instabilities is reduced to that of
diagonalizing matrices $S_{{\delta }}\left( {I,K}\right),$ with $%
{\delta }$ fixed. The eigenvectors determine the order parameter
characterizing the instability.

For charge density wave we define ${\cal R}_{{\delta}}\left( {I,K}%
\right) \equiv D_{{\delta }}^{c}\left( {I,K}\right) ,$ with the
order parameter ${\cal O}_{CDW,{\delta }}\left( {I}\right) =\frac{1}{%
\sqrt{2}}\sum_{\tau}\,R_{{I+\delta},\tau }^{+}\,L_{{I},\tau}\,$\
\ $\equiv \frac{1}{\sqrt{2}}\sum_{\tau}\,R_{{M+Q},\tau }^{+}\,L_{{ %
M-Q},\tau }\;$(i.e., ${M=I+}\frac{{ \delta }}{2}$ and $\,{ Q=}%
\frac{{ \delta }}{2}$). The last definition stresses that
interactions connect only order parameters with equal
particle-hole center of mass, ${%
M.}$ Thus, the problem of diagonalizing $D_{{\delta }}^{c}\left( {I,K%
}\right) $ is once again simplified as there is a fixed parameter. Finally
for spin density waves we define similarly ${\cal O}_{SDW,{\delta }%
}\left( {I}\right) =$\ $\frac{1}{\sqrt{2}}%
\sum_{\tau }\,\tau \,R_{{M+Q},\tau }^{+}\,L_{{M-Q},\tau }.$

$\strut $With these definitions it is straightforward to obtain RG equations
for the response functions\cite{Solyom79}$^{,}$\cite{Douçot}, of the form $%
\partial \ln \overline{{\cal R}}_{{\delta }}\left( {I,K}\right)
/\partial \ln D=\upsilon $ with $\overline{{\cal R}}_{{\delta
}}\left( {I,K}\right) $= $\pi \left( v_{{ I}+{ \delta }}+v_{{
I}}\right)
\partial {\cal R}_{{ \delta }}\left( { I,K}\right) /\partial \ln
\omega $. More precisely:

\begin{eqnarray}
\upsilon _{SSC} &=&-\frac{C_{{ I-K}}^{c}\left( { K},{ K}+{ %
\delta }\right) -3C_{{ I-K}}^{s}\left( { K},{ K}+{ \delta }%
\right) }{ \left( v_{{ K}+{ \delta }}+v_{{ K}}\right) }
\label{eSSC} \\
\upsilon _{TSC} &=&-\frac{C_{{ I-K}}^{c}\left( { K},{ K}+{ %
\delta }\right) +C_{{ I-K}}^{s}\left( { K},{ K}+{ \delta }%
\right) }{\left( v_{{ K}+{ \delta }}+v_{{ K}}\right) }
\label{eTSC} \\ \upsilon _{CDW} &=&\frac{C_{{ I-K}}^{c}\left( {
K,I+\delta }\right)
+3C_{{ I-K}}^{s}\left( { K,I+\delta }\right) }{\left( v_{{ %
2I-K+\delta }}+v_{{ K}}\right) }  \label{eCDW} \\ \upsilon _{SDW}
&=&\frac{C_{{ I-K}}^{c}\left( { K,I+\delta }\right)
-C_{{ I-K}}^{s}\left( { K,I+\delta }\right) \,\,}{\left( v_{%
{ 2I-K+\delta }}+v_{{ K}}\right) }  \label{eSDW}
\end{eqnarray}

\strut In this work we are concerned with the ordering of the
electronic system. As a general trend, the effective couplings in
the one-loop approximation diverge for a finite value of the
reduced cut-off. This is naturally interpreted as the signature of
a low temperature instability towards an ordered state. However
this divergency of the effective couplings also signals a
breakdown of the one-loop approximation. This difficulty may be
overcome by noting that in general ratios of any two coupling
constants reach finite asymptotic values as the instability is
approached\cite{Balents}. In this work we have thus emphasized the
study of these asymptotic flows in the space of {\em directions
}for the effective coupling vector. Representing the $C_{{\delta
}}({I,J})$ by a generic $C_{i},$ this
can be done by finding the flow of non-diverging normalized couplings $%
c_{i}\equiv C_{i}/{\cal N},$ with ${\cal N=}\sqrt{\sum_{j}\left(
C_{j}\right) ^{2}}$ (the sum is over all the couplings). From the RG
equations $\left( \ref{RG1}\right) $ and $\left( \ref{RG2}\right) $ having
the form  $\partial C_{i}/\partial \ln D=$ $A_{ijk}\,C_{j}C_{k}$ we obtain
the evolution for the normalized couplings: $dc_{i}/du=\sum_{jkl}A_{jkl}%
\,c_{k}$\thinspace $c_{l}\left( \delta _{ij}-c_{i}c_{j}\right) .$
We have introduced a new scale parameter $u$ related to $D$ by
$du=-{\cal N}$ $d\ln D $. In this parametrization the magnitude
${\cal N}$ of the coupling vector diverges only as $u\rightarrow
\infty .$ From the flow as a function of $u$,
we can recover the more physical variable $D$ since $d\ln {\cal N}/du$= $%
A_{ikl}\,c_{i}\,c_{k}\,c_{l}$ which yields ${\cal N}\left( u\right) $ and
therefore $D\left( u\right) $. A similar procedure can be applied to the
computation of response functions, that we generically represent by $%
\overline{{\cal R}}_{i}$. Their RG equations have the form $d\ln \overline{%
{\cal R}}_{i}/d\ln D=B_{ik}\,C_{k}$, which becomes $d\ln \overline{{\cal R}}%
_{i}/du=B_{ik}\,c_{k}$. As $u$ goes to infinity the r.h.s. reaches a finite
value. This is easily seen to imply a divergency of the form ${\cal R}%
_{i}\sim [\,\ln (D/D^{*})\,]^{-\alpha _{i}}$, where $D^{*}$ is the scale at
which the instability occurs. The exponents $\alpha _{i}$ are completely
determined by the fixed point values $c_{i}^{*}$ of the normalized
couplings.

This procedure provides a controlled framework to study the complete RG flow
(not restricted to the one-loop approximation) provided the bare couplings
are small enough. Certainly we expect the one-loop approach to break down if $%
{\cal N}$ is of the order unity. For a given choice of initial couplings
with norm ${\cal N}_{0}$, this yields a maximal value of $u$ ($u_{\max }$)
beyond which our approximation is no longer reliable. The main point is that
$u_{\max }$ can be made arbitrarily large provided ${\cal N}_{0}\,$is small
enough. In general we expect  $u_{\max }\sim \ln (1/{\cal N}_{0})$. In some
applications we are also interested in a finite an not infinitesimal ${\cal N%
}_{0}$ regime. Higher order corrections destroy the fixed directions found
in the one-loop approximation. However, we conjecture those become fixed
points in the usual sense and that the topology of the one-loop flow pattern
is preserved.

\mathstrut
%%%%%%%%%%%%%%%%%%%%%
%      figure       %
%%%%%%%%%%%%%%%%%%%%%
\begin{figure}[tbp]
     \begin{center}
       \leavevmode
       \hbox{%
       \epsfxsize 2.4in \epsfbox{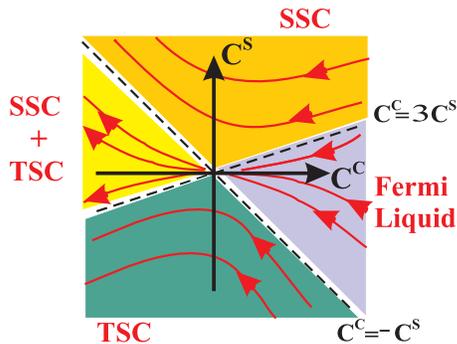}}
       \end{center}
\caption{ Phase diagram for a two dimensional isotropic electronic
system with a circular Fermi surface.}
\end{figure}
%%%%%%%%%%%%%%%%%%%%%
%      end          %
%%%%%%%%%%%%%%%%%%%%%

As a first example of application of the present formalism we
consider a non-nested Fermi surface. The spinless case is
thoroughly discussed in the
review paper by Shankar\cite{Shankar}. Phase space restrictions set in $%
\left( \ref{RG1}\text{-}\ref{RG2}\right) $ through the $\delta $
functions, imply that only couplings of the form $C_{{ \delta }}({
I,I})$ are renormalized, due to particle-particle corrections (BCS
channel). Divergence of the flow is then related to
superconductivity. Indeed, inspired from (\ref{eSSC}-\ref{eTSC}),
the RG flow equations can be quite
simply written in terms of singlet $C_{{ \delta }}^{S}({ I})\equiv C_{%
{ \delta }}^{c}({ I,I})-3C_{{ \delta }}^{s}({ I,I})$ and triplet
couplings$\,C_{{ \delta }}^{T}({ I})\equiv C_{{ \delta }}^{c}({ %
I,I})+C_{{ \delta }}^{s}({ I,I})$:
\begin{equation}
\partial C_{{ \delta }}^{T}({ I})=\alpha \left( C_{{ \delta }}^{T}(%
{ I})\right) ^{2}\;\;\;\;\;\;\partial C_{{ \delta }}^{S}({ I}%
)=\alpha \left( C_{{ \delta }}^{S}({ I})\right) ^{2} \label{sh}
\end{equation}
where $\alpha $ is a Fermi velocity dependent constant. The phase
diagram is
shown in Figure 2, for a system with initial couplings $C_{{ \delta }%
}^{c}({ I,I})=C^{c}$ and $C_{{ \delta }}^{s}({ I,I})=C^{s}$. The
only region where the flow does not diverge corresponds to a Fermi
Liquid phase\cite{Shankar}. In the other regions, the RG couplings
diverge at the energy scale that we associate with the
superconducting critical temperature, $T_{c}$.
%%%%%%%%%%%%%%%%%%%%%
%      figure       %
%%%%%%%%%%%%%%%%%%%%%
\begin{figure}[tbp]
       \begin{center}
       \leavevmode
       \hbox{%
       \epsfxsize 2.0in \epsfbox{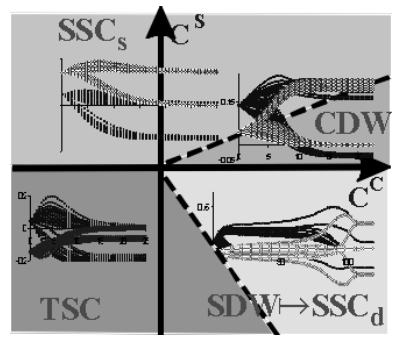}}
       \end{center}
       \caption{Phase diagram for a nested system: the flows of the couplings in each
       phase are inserted.}
\end{figure}
%%%%%%%%%%%%%%%%%%%%%
%      end          %
%%%%%%%%%%%%%%%%%%%%%

This picture changes when we introduce nesting. As already
discussed \cite{Zheleznyak,Fernao} flat regions dramatically
enhance the number of low energy couplings to be considered. The
occurrence of log singularities in both the particle-particle and
the particle-hole channels induce an intricate interference
mechanism, leading to the so-called {\em parquet regime}. In
Figure 3, we show the new phase diagram, together with typical
flow patterns (using the $u$ variable, and normalized couplings)
for each of the different phases. These results have been obtained
for a partially nested Fermi surface with $N=6$ patches along each
flat segment and $M=0$ along each curved arc. Note that increasing
the number of patches doesn't change significantly the
phase-diagram.

The modifications due to nesting are most obvious in the initial
coupling domain corresponding to the Fermi liquid regime for the
circular Fermi surface. For positive bare spin coupling, we obtain
a CDW phase, as a mean field analysis predicts. The most striking
result appears for repulsive couplings in the charge sector and
moderate negative spin couplings. With this parametrization, the
simple repulsive Hubbard model falls in this region. The flow
pattern from Figure 3 clearly shows a two-step evolution as $u$ is
increased. Before entering the final asymptotic regime, the system
spends a large $u$-time in the proximity of an unstable fixed
point (for the normalized couplings). In this intermediate phase,
the most diverging response is the SDW susceptibility. The
corresponding order parameter has the approximate form
$R^{+}_{N}L_{1}+R^{+}_{N-1}L_{2}+...+h.c$ so the most favorable
nesting vector is perpendicular to the flat regions. Couplings
involving incoming and outgoing electrons close to the endpoints
of flat regions are the most enhanced. This intermediate fixed
point corresponds very likely to the one found by Zheleznyak et al
\cite{Zheleznyak} who worked with the original un-normalized
couplings. Our approach yields the remarkable result that the
system is finally attracted towards a stable low energy d-wave
superconductor ($SSC_{d}$) fixed point. Again, the largest
couplings involve the endpoints of the flat regions, and from the
superconducting response function, the gap exhibits a strong
anisotropy. It vanishes at the center of flat regions and reaches
its maximal values at their extremities. The dramatic role played
by these extremal regions suggests some similarity with the
Van-Hove singularity scenario studied by several groups
\cite{Zanchi,Honerkamp,Guinea2000}. However, we haven't introduced
any singularity in the Fermi velocity. Our model also ignores
Umklapp processes, which are expected to play a crucial role in
stabilizing a commensurate SDW (N\'eel state), and in describing
the proximity to a Mott insulator \cite{Honerkamp}.

It is important to specify the energy scales associated to both
fixed points. Assuming the bare bandwidth is $D_{0}$, the
correspondence between the running scale $D$ and the fictitious
time $u$ takes the form: $ln(D_{0}/D)=h(u)/ {\cal N}_{0}$, where
${\cal N}_{0}$ is the length of the initial coupling vector, and
$h(u)$ is a function depending on the direction of this vector. In
general, $h$ appears to be an increasing function of $u$ which
reaches a finite limit $h(\infty)$ as $u$ becomes infinite. This
defines the instability scale $D_{c}$ below which the system is
superconducting. Similarly, we define a scale $D_{1}$ for which
the distance of the normalized coupling vector to the intermediate
fixed point is minimal. Let us note by $u_{1}$ the corresponding
value of $u$. We obtain:
$D_{1}/D_{0}=(D_{c}/D_{0})^{h(u_{1})/h({\infty})}$, which holds
for a chosen direction of the initial coupling vector. $D_{c}$ and
$D_{1}$ are considered here to be functions of ${\cal N}_{0}$. In
practice, we have found that $h(u_{1})/h({\infty})$ is very close
to unity. As an order of magnitude, if $D_{c} \simeq 100K$ and
$D_{0} \simeq 5000K$, $D_{1}-D_{c} \simeq 1K$. However, SDW
correlations begin to build up at a much higher energy scale, as
can be seen in Figure 4. We may for instance evaluate the scale
$D^{*}$ at which the SDW exponent reaches 99 percent of its
maximal value.

\begin{center}
\begin{tabular}{|c|c|c|c|c|c|c|}
\hline \ \ \ \ $N$ \ \ \ \ & \ \ \ $3  \ \ \ $ & \ \ $6  \ \ $ & \
\ $8 \ \ $ & $ \ \ \ 10 \ \ \ \ $ & $ \ \ \ 12 \ \ \ $ & \ \ \ $
14 \  $ \
\\ \hline $T_{c}\left( K\right) $ & $0$ & $0.0$ & $0.2$ & $15$ &
$82$ & $150$ \\ \hline $T^{*}\left(K\right) $ & $0$ & $0.0$ & $9$
& $208$ & $736$ & $1066$ \\ \hline \multicolumn{7}{l}{{\small
Table 1: System with N+M=14 and normalized couplings.}}
\end{tabular}
\end{center}

From the table above, we conclude that there can be a large
temperature range for which a SDW enhancement is observed above
the superconducting instability. Both temperature scales are seen
to decrease as the length of nested segments is reduced. This
reduction is accompanied by an increase in the minimal distance to
the unstable fixed point along the flow. This leads to a less
pronounced intermediate regime, smaller exponents for the SDW
response, and consequently a smaller magnitude of SDW
correlations.
%%%%%%%%%%%%%%%%%%%%%
%      figure       %
%%%%%%%%%%%%%%%%%%%%%
\begin{figure}[tbp]
       \begin{center}
       \leavevmode
       \hbox{%
       \epsfxsize 2.4in \epsfbox{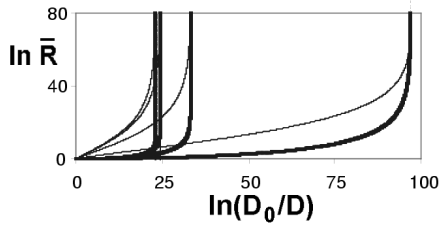}}
       \end{center}
       \caption{Flow of the dominant SDW (thin) and SSCd (thick) response functions
       for systems with N+M=14, but different nesting qualities (N=14,12,8,3).
       Better nested systems develop instabilities at higher energy scales. SSCd
       response functions become the most diverging responses at the final energy
       scale, associated with $T_{c}$.}
\end{figure}
%%%%%%%%%%%%%%%%%%%%%
%      end          %
%%%%%%%%%%%%%%%%%%%%%

In summary, we have shown that precursor effects are predicted by
a controlled weak coupling RG approach to fermionic systems. The
flow approaches an unstable fixed point for systems with repulsive
interactions, before being attracted to the final stable
superconducting fixed point controlling the low temperature
physics of the system. SDW response functions are enhanced and
dominate near the intermediate fixed point, before the d-wave
superconducting response finally diverges. We believe that Umklapp
processes are likely to strengthen this picture. Close to
half-filling, they are not expected to compete with SDW ordering,
but with superconductivity, thereby reducing $T_{c}$ without
affecting the cross-over energy scale. In spite of the weak
coupling limit chosen here, we observe striking similarities with
the pseudogap behavior of the HTC. This suggests that the strength
of interactions in real materials may compensate the deviations of
the observed Fermi surface from perfect nesting.

The authors acknowledge discussions with D. Zanchi, K. Le Hur and
R. Dias. This work received traveling support from ICCTI/CNRS,
Proc.423. FVA also benefited from the MCT Praxis XXI Grant No.
2/2.1/Fis/302/94.

\end{document}